\documentclass[a4paper,
              ]{jacow}
\makeatletter%
	\ifboolexpr{bool{xetex}}
	 {\renewcommand{\Gin@extensions}{.pdf,%
	                    .png,.jpg,.bmp,.pict,.tif,.psd,.mac,.sga,.tga,.gif,%
	                    .eps,.ps,%
	                    }}{}
\makeatother

\ifboolexpr{bool{xetex} or bool{luatex}} % test for XeTeX/LuaTeX
 {}                                      % input encoding is utf8 by default
 {\usepackage[utf8]{inputenc}}           % switch to utf8

\usepackage[USenglish]{babel}			 

\usepackage[final]{pdfpages}
\usepackage{multirow}
\usepackage{ragged2e}
\usepackage{hyperref,graphicx,adjustbox}

\ifboolexpr{bool{jacowbiblatex}}%
 {%
  \addbibresource{jacow-test.bib}
  \addbibresource{biblatex-examples.bib}
 }{}
\begin{document}

\title{Experiments on Electron Cooling and Intense Space-Charge at IOTA}
\author{N. Banerjee\thanks{nilanjan@fnal.gov}, G. Stancari, Fermilab, Batavia, Illinois, USA \\
M.K. Bossard, J. Brandt, Y-K. Kim, The University of Chicago, Illinois, USA \\
S. Nagaitsev, Thomas Jefferson National Accelerator Facility, Newport News, Virginia, USA}
		
\maketitle
\begin{abstract}
    The Integrable Optics Test Accelerator at Fermilab will explore beam dynamics in a ring with intense space-charge using 2.5~MeV proton beams with an incoherent tune shift approaching -0.5. We will use this machine to explore the interplay between electron cooling, intense space-charge, and coherent instabilities. In this contribution, we describe the machine setup including the design of the electron cooler and the lattice, list specific experiments and discuss the results of numerical simulations which include the effects of electron cooling and transverse space-charge forces.
\end{abstract}

\section{Introduction}
\begin{figure}
    \centering
    \includegraphics[width=\columnwidth]{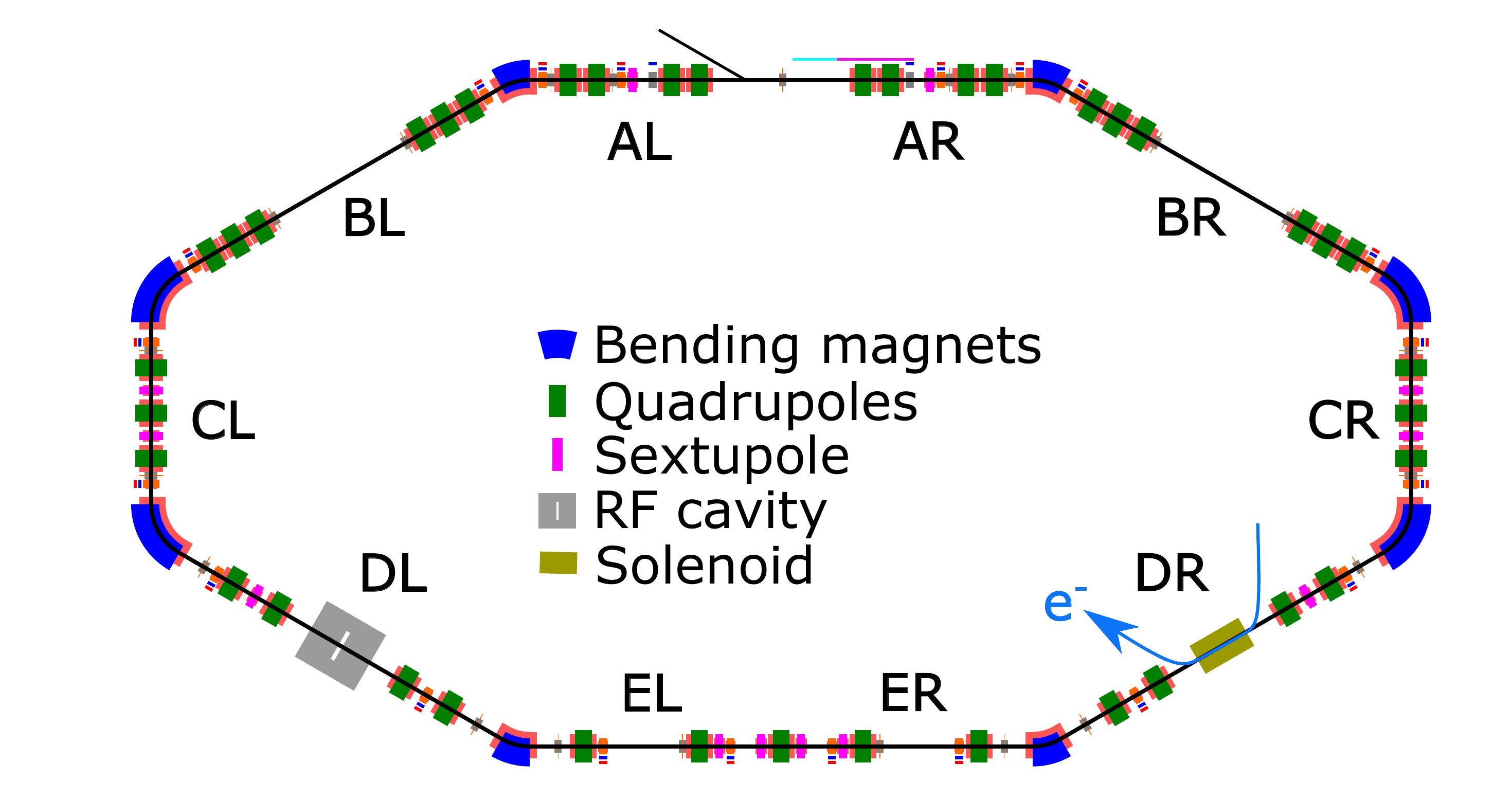}
    \caption{Layout of the Integrable Optics Test Accelerator\cite{Antipov2017} without special non-linear magnets. The beam moves clockwise. The blue arrow indicates the propagation of electrons in the cooler situated in section DR.}
    \label{fig:iota}
\end{figure}

The grand challenges facing the accelerator and beam physics community include creating and sustaining beams with intensity and phase-space density an order of magnitude higher than what is achievable today.\cite{Blazey2023} In the realm of hadron storage rings, such gains are crucial for future proton drivers for neutrino generation\cite{Garoby2016}, neutron sources\cite{Galambos2020} and muon colliders\cite{Boscolo2019}, along with heavy ion colliders for particle and nuclear physics. Specifically at Fermilab, the Accelerator Complex Evolution plan\cite{Ainsworth2023}, which aims to provide substantially more protons on target when compared to PIP-II\cite{Lebedev2017}, requires the replacement of the Fermilab Booster\cite{Eldred2021} synchrotron. One class of options involves constructing a rapid cycling synchrotron which will operate at high intensity and high space-charge tune shift to sustain 2.4~MW beam power on the Long-Baseline Neutrino Facility (LBNF) target. The Integrable Optics Test Accelerator (IOTA)\cite{Antipov2017}, displayed in Fig.~\ref{fig:iota}, was constructed at Fermilab as part of the R\&D program toward achieving multi-megawatt proton beams. Research at IOTA includes Non-linear Integrable Optics\cite{Valishev2021, Wieland2023}, coherent instabilities, beam cooling\cite{Jarvis2022}, electron lenses\cite{Stancari2021}, and more\cite{Lobach2021, Romanov2021}. IOTA was designed to operate both with 150~MeV electrons, which we have been using until now, and also 2.5~MeV protons, whose injector we are building. The proton program in IOTA is designed for experiments up to an incoherent tune shift of -0.5 to explore methods of improving the stability of bright and intense hadron beams in synchrotrons and storage rings.\\

The primary mechanism through which space-charge affects beam dynamics in hadron storage rings is through betatron resonance excitations\cite{Hofmann2015} due to incoherent tune shifts of the particles as a function of position inside the bunch. In practice, periodic resonance crossing\cite{Franchetti2003} of particles undergoing synchrotron motion in bunched beams lead to emittance growth and beam loss, thus limiting the maximum phase-space density which can be sustained in a ring. Studying these effects in high-energy hadron rings is complicated since the tune spread generated by space-charge dynamically changes the phase-space distribution of the bunch, which in turn influences the tune shift itself. Consequently, by using an electron cooler to enforce an equilibrium phase-space distribution in the bunch core, we can measure halo formation and beam loss in the presence of almost constant space-charge forces. Past experiments using electron cooling, seeking to maximize the Laslett tune shift of accumulated beam, found an upper limit of $|\Delta\nu_\perp| \sim 0.1 - 0.2$.\cite{Nagaitsev1995, Steck2000} Using magnetized electron cooling of 2.5~MeV proton beams in IOTA, we seek to maximize the phase-space density of stored beam, measure beam loss and characterize the distribution.\\

While space charge restricts the maximum phase-space density of the beam, the onset of coherent excitations, such as the Transverse Mode Coupling Instability (TMCI), constrains the maximum current. In contrast, incoherent particle motion in the bunch with different betatron tunes results in Landau damping, which naturally restricts the growth of these instabilities. We can exploit this effect by providing non-linear focusing to the beam using octupoles\cite{Antipov2021} or electron lenses\cite{Shiltsev2017}, and this has been instrumental in boosting the beam current in many accelerators. However, strong non-linear focusing also restricts the dynamic aperture, which constrains the maximum phase-space volume occupied by the bunch. We will explore Non-linear Integrable Optics (NIO)\cite{Danilov2010, Danilov1997} at IOTA, which can produce adjustable tune spread while ensuring stable single-particle dynamics where theoretically, the aperture is solely limited by the vacuum chamber. Besides dedicated non-linear optics, space-charge fields also generally provide non-linear focusing and, hence, can serve as a source of incoherent tune spread. Historically, this was believed to have a damping effect on TMCI, where increasing space-charge tune shift increases the TMCI current threshold. But recent analyses\cite{Burov2019, Burov2018, Buffat2021} indicate a complex interplay between space charge and wakefields. We will measure instability growth rate and damping effects using 2.5~MeV protons at IOTA with electron cooling serving as the knob for the space-charge tune shift, a digital wake-building feedback system\cite{Ainsworth2021} providing variable impedance, and non-linear integrable optics contributing amplitude-dependent tune spread.\\

The electron cooler will be realized as a specific configuration of the IOTA electron lens\cite{Stancari2021} in section DR as illustrated in Fig.~\ref{fig:iota}. The electron lens consists of a low-energy (<10~keV) magnetically confined electron beam which interacts with the recirculating beam and can be used for non-linear focusing, space-charge compensation, and electron cooling. The primary components of the setup include a thermionic electron source, a collector, and a magnetic system consisting of a main solenoid for the interaction region and transfer solenoids and toroids to guide the beam. The design of the main solenoid is constrained by the maximum axial field requirement for applying the lens to 150~MeV electrons and the strict field quality requirement dictated by electron cooling of 2.5~MeV protons. Diagnostics specific to analyzing electron cooling performance include a neutral hydrogen monitor to non-invasively measure the equilibrium proton beam profile during cooling and a cyclotron radiation emission monitor\cite{Asner2015, Habfast1987} to measure electron density and temperature. While we already have many components from the decommissioned Tevatron electron lens, we are finalizing the design of the vacuum chambers and the transport system to meet the stringent requirements of electron cooling.\\

In the next section, we define the specifications for the 2.5~MeV proton beam in IOTA and the corresponding electron cooler. Then, we outline specific experiments which probe the frontier of phase-space density and intensity of the proton beam.\\

\begin{table}
    \caption{Proton beam parameters in IOTA.}
    \label{tab:protonops}
    \centering
     \begin{adjustbox}{max width=0.95\columnwidth}
    \begin{tabular}{p{\dimexpr 0.4\linewidth-2\tabcolsep}
                    p{\dimexpr 0.25\linewidth-2\tabcolsep}
                    p{\dimexpr 0.24\linewidth-2\tabcolsep}
                    p{\dimexpr 0.11\linewidth-2\tabcolsep}}
        \hline
        \textbf{Parameter} & \multicolumn{2}{c}{\textbf{Value}} & \textbf{Unit} \\
        \hline
        Kinetic energy ($K_b$) & \multicolumn{2}{c}{2.5} & MeV \\
        Emittances ($\epsilon_{x,y}$) & \multicolumn{2}{c}{4.3, 3.0} & $\mu$m \\
        Momentum spread ($\sigma_p/p$) & \multicolumn{2}{c}{$1.32\times10^{-3}$} & \\
        \hline
        & \textbf{Coasting} & \textbf{Bunched} &\\
        \hline
        Number of bunches & - & 4 & \\
        Bunch length ($\sigma_s$) & - & 0.79 & m \\
        Beam current ($I_b$) & 5.79 & 1.15 & mA \\
        Bunch charge ($q_b$) & 10.6 & 0.52 & nC \\
        Tune shifts ($|\Delta \nu_{x,y}|$) & \multicolumn{2}{c}{0.33, 0.50} & \\
        \hline
        $\tau_{\text{IBS,x,y,z}}$ & 10.2, 2.62, 301 & 14.4, 3.70, 424 & s \\
        \hline
    \end{tabular}
\end{adjustbox}
\end{table}

\section{Electron Cooling with 2.5~MeV Protons}
The proton program is designed for experiments with both coasting and bunched beams, with space-charge tune shifts approaching -0.5. The 2.5~MeV ($pc \approx 70$~MeV) beam energy has the double advantage of large tune shifts being achievable using modest bunch charges, as shown in Table~\ref{tab:protonops}, and simultaneously being able to produce almost no intrinsic impedance, enabling us to disentangle intensity effects. The proton injector\cite{Edstrom2023} will produce the required beam parameters at a repetition rate of 1~Hz, and the beam will be injected in a single turn into section A. The IOTA rf system\cite{Bruhaug2016} located in section DL contains two broadband resonators driving independent accelerating gaps, one of which operates at 2.2~MHz for bunched proton beam operations at $h=4$. The placement of linear focusing elements in IOTA preserves mirror symmetry about the vertical centerline of the layout presented in Fig.~\ref{fig:iota}. However, we can independently control all magnet strengths, allowing us to set up lattices with a wide range of tunes, transverse coupling, momentum compaction, along with special configurations required for NIO. The lattice\cite{Banerjee2023} for electron cooling experiments enforces zero dispersion and transverse betatron matching at the cooling solenoid and features adjustable main solenoid strength, transverse tunes, and linear coupling. The drawback of low-energy operation is the large emittance growth and loss rates driven by Intra Beam Scattering (IBS) and residual gas scattering. Consequently, electron cooling is necessary to extend the lifetime of the proton beam and maintain equilibrium conditions.\\

\begin{figure}
    \centering
    \includegraphics[width=\columnwidth]{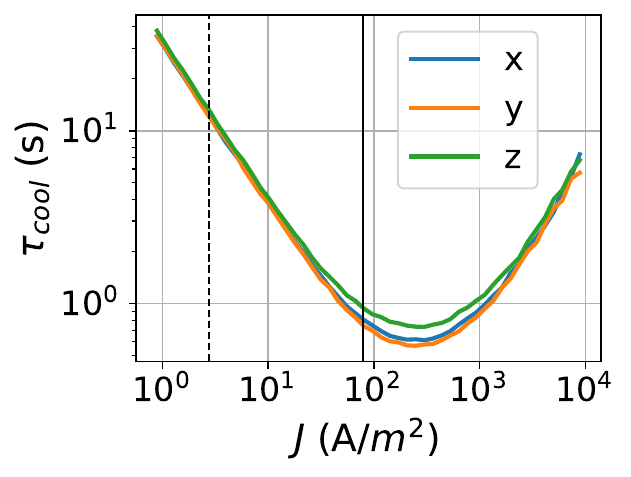}
    \caption{Cooling times of a Gaussian bunched beam as functions of electron current density in the presence of axial magnetic field of 0.1~T without any errors. The vertical lines correspond to the electron beam configurations listed in Table~\ref{tab:ecooler}.}
    \label{fig:tcool}
\end{figure}

\begin{table}
    \caption{Electron cooler parameters.}
    \label{tab:ecooler}
    \centering
    \begin{adjustbox}{max width=0.95\columnwidth}
    \begin{tabular}{p{\dimexpr 0.4\linewidth-2\tabcolsep}
                    p{\dimexpr 0.25\linewidth-2\tabcolsep}
                    p{\dimexpr 0.25\linewidth-2\tabcolsep}
                    p{\dimexpr 0.11\linewidth-2\tabcolsep}}
        \hline
        \textbf{Parameter} & \multicolumn{2}{c}{\textbf{Values}} & \textbf{Unit} \\
        \hline
        \multicolumn{4}{c}{\textbf{Proton parameters}} \\
        \hline
        RMS Size ($\sigma_{b,x,y}$) & \multicolumn{2}{c}{4.43, 3.70} & mm\\
        \hline
        \multicolumn{4}{c}{\textbf{Main solenoid parameters}} \\
        \hline
        Magnetic field ($B_\parallel$) & \multicolumn{2}{c}{0.1 - 0.5} & T \\
        Length ($l_\text{cooler}$) & \multicolumn{2}{c}{0.7} & m \\
        Flatness ($\max B_\perp/B_\parallel$) & \multicolumn{2}{c}{$2\times10^{-4}$} & \\
        \hline
        \multicolumn{4}{c}{\textbf{Electron parameters}} \\
        \hline
        Kinetic energy ($K_e$) & \multicolumn{2}{c}{1.36} & keV \\
        Temporal Profile & \multicolumn{2}{c}{DC or pulsed} & \\
        Transverse Profile & \multicolumn{2}{c}{Flat} & \\
        Source temp. ($T_\text{cath}$) & \multicolumn{2}{c}{1400} & K \\
        \hline
        Current ($I_e$) & 1.7 & 80 & mA \\
        Radius ($a$) & 14 & 18 & mm \\
        $\tau_\text{cool,x,y,s}$ & 12, 12, 13 & 0.79, 0.74, 0.94 & s\\
        \hline
    \end{tabular}
\end{adjustbox}
\end{table}

The goal of the electron cooler is to compensate for various emittance growth mechanisms and provide us with a knob to tune the equilibrium phase-space distribution for a range of proton bunch charges and energies. This translates to requiring a range of cooling time scales between 1 to 10 seconds. We opt for magnetized electron cooling with main solenoid strengths ranging between 0.1-0.5~T to boost cooling rates and provide transverse confinement to the low-energy electron beam. We calculate the cooling times of a bunched proton beam with a Gaussian distribution in all directions, assuming that (1) the velocity at the center of the electron beam matches with the average velocity of the protons, (2) the transverse distribution of the protons is matched to the beta function of the solenoid, and (3) the magnetic field at the cathode equals that at the main solenoid. Figure~\ref{fig:tcool} shows the results of the cooling time calculations using the code JSPEC\cite{Zhang2021} as a function of current density of the electrons. The cooling rate increases linearly with current density for low electron currents, reaches a maximum, and then declines for high currents. This can be explained in terms of the relative contributions of space-charge, electron beam energy jitter due to power supply fluctuations, and magnetic field errors to the effective velocity in the Parkhomchuk model.\cite{Parkhomchuk2000} The cooling rate increases linearly at low electron currents ($J\lesssim 20$~A/m\textsuperscript{2}), since the effective velocity is dominated by power supply jitter and magnetic field errors, both of which are independent of current. At higher currents, the relative motion of electrons in the beam frame, driven by space-charge forces, contributes more to the effective velocity as intensity rises. This leads to diminishing cooling rates even at larger number densities of electrons. In a practical realization, the maximum electron density will be limited by the highest betatron tune shift $\Delta \nu_{\perp,\mathrm{cooler}}$, which can be accommodated by the lattice while retaining linear stability. In our lattice, the limit of stability corresponds to $\Delta \nu_{\perp,\mathrm{cooler}} \sim 0.1$, which requires $J\lesssim10^3$~A/m\textsuperscript{2}. We chose two thermionic electron source designs\cite{Bossard2023} listed in Table~\ref{tab:ecooler} with an order of magnitude difference in current density to cool protons over a range of bunch charges.\\

\begin{figure*}
    \centering
    \includegraphics[width=0.9\textwidth]{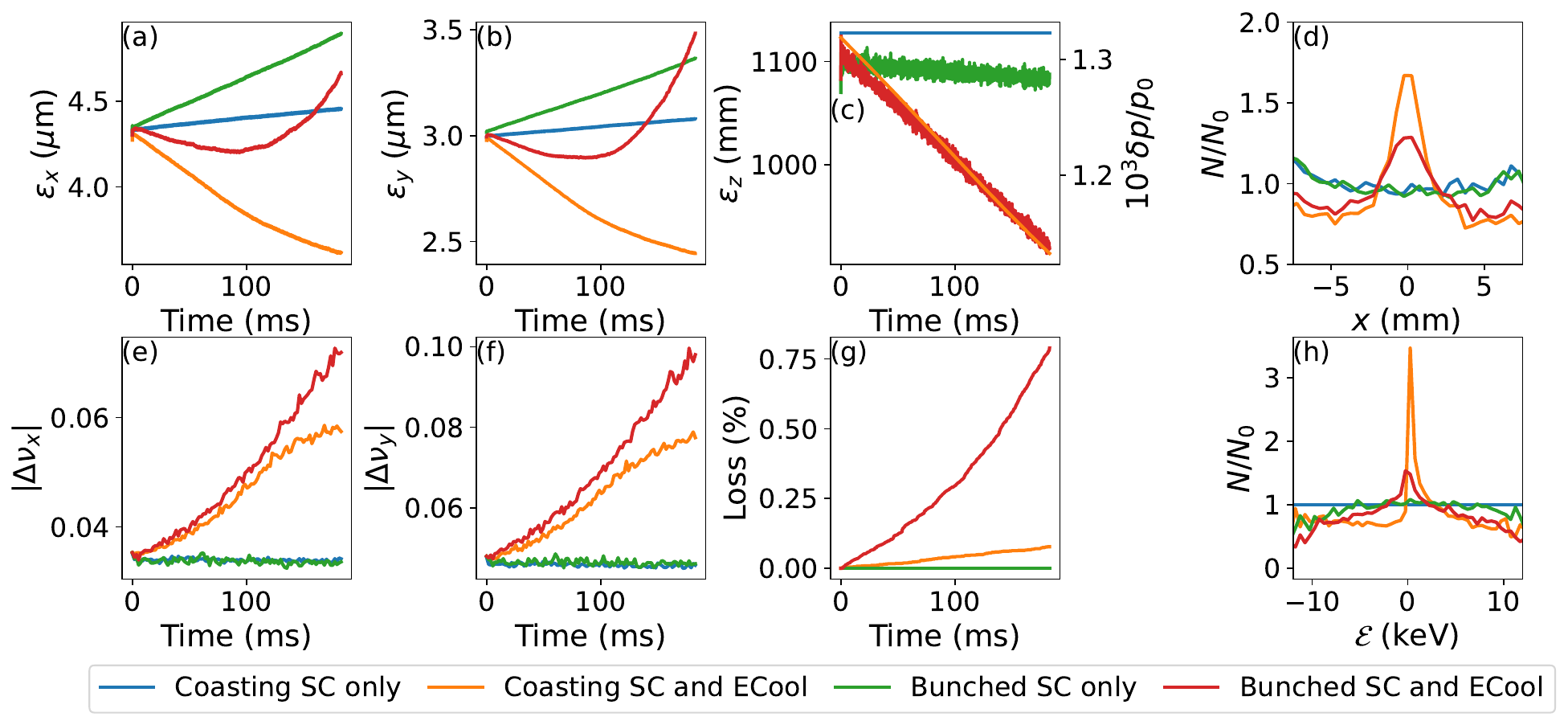}
    \caption{Evolution of beam properties in PyORBIT simulations of 2.5~MeV protons in IOTA.(a) Horizontal rms emittance. (b) Vertical rms emittance. (c) Longitudinal rms emittance. (d) Change in horizontal particle distribution $N/N_0 \equiv \{\mathrm{d}N(t=183\,\mathrm{ms})/\mathrm{d}x\}/\{\mathrm{d}N(t=0)/\mathrm{d}x\}$ (e) Horizontal incoherent tune shift in the bunch core. (f) Vertical incoherent tune shift in the bunch core. (g) Percentage of beam lost. (h) Change in energy distribution $N/N_0 \equiv \{\mathrm{d}N(t=183\,\mathrm{ms})/\mathrm{d}\mathcal{E}\}/\{\mathrm{d}N(t=0)/\mathrm{d}\mathcal{E}\}$. Note that the slight reduction in longitudinal emittance for the bunched beam with space-charge only (green line in panel (c)) is due to a slight mismatch of the injected distribution.}
    \label{fig:tuneshiftsandloss}
\end{figure*}

We model the interplay of transverse space-charge and magnetized electron cooling using the Particle-in-Cell code \texttt{PyORBIT}\cite{Shishlo2015}. Our implementation\cite{Banerjee2021} uses the \texttt{SpaceCharge2p5D} model in \texttt{PyORBIT} and applies multiple thin-lens kicks to the protons, which sum the contribution of the cooling and the focusing force from the electron beam as they propagate through the solenoid. We conduct bunched and coasting beam simulations to demonstrate this model, injecting a Gaussian distribution with $|\Delta\nu_y|=0.05$ and allowing it to evolve in IOTA with and without electron cooling for 100,000 turns (183~ms). Figure~\ref{fig:tuneshiftsandloss} shows the rms emittances, transverse tune shifts at the core, and the beam losses as functions of time. In the absence of cooling, the transverse rms emittances for both coasting (blue) and bunched (green) beams rise linearly with a time scale of 1-2~seconds, which gradually reduces the incoherent tune shift of the beam. With cooling, the rms emittances of the coasting beam (orange) continue shrinking throughout the duration of the simulation, but for the bunched beam (red), they reach a minimum at 100~ms and then start expanding. In contrast, the transverse incoherent tune shifts at the bunch core increase monotonically as a function of time, where the vertical tune shift reaches 0.1 at the end of the simulation for the bunched beam and 0.08 for the coasting beam. This is consistent with the observation of growing phase-space density at the bunch core as seen in panel (d) in Fig.~\ref{fig:tuneshiftsandloss}. Panel (g) shows substantial beam loss for the bunched beam with electron cooling in a relatively short period of time, which is much less for the coasting beam and completely absent in the simulations with space charge only. Periodic resonance crossing of small-amplitude particles in the bunch may explain the formation of tails and the subsequent beam loss, but more simulations are required to verify this. We will use these simulation models to optimize our experiments at IOTA and gain insight into the observations.\\

\section{Beam Physics Experiments using Electron Cooling}
\begin{figure*}
    \centering
    \includegraphics[width=0.8\textwidth]{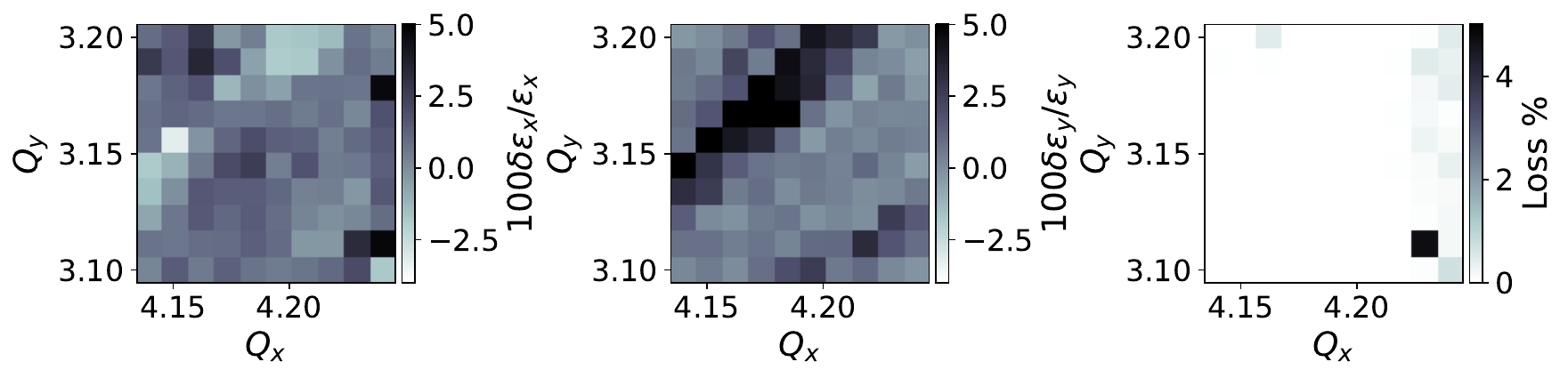}
    \caption{Simulated change of rms emittance and beam loss over 2500~turns as a function of bare-lattice tune for a bunched beam with an initial vertical space-charge tune shift of 0.05 injected into the IOTA lattice with no errors.}
    \label{fig:tunescan}
\end{figure*}

Space-charge-induced periodic resonance crossing is regarded as the mechanism behind emittance growth and the loss of bunched beams, hence constraining the maximum bunch charge\cite{Oeftiger2022} sustainable within a given emittance growth and beam loss budget. The observed maximum limit ($|\Delta\nu_\perp| \sim 0.1 - 0.2$) on incoherent tune shift of accumulated beam in ion rings with electron cooling is presumably due to the same mechanism. We want to measure the maximum proton bunch charge sustainable in a storage ring with electron cooling for a given duration and a fixed loss budget. We will use the typical methods, including optimization of the working point of the bare lattice and compensation of resonance driving terms\cite{Asvesta2022} using sextupoles and octupoles to minimize beam loss. Figure~\ref{fig:tunescan} demonstrates results from a simulated low-resolution ($\Delta Q = 0.01$) scan of bare-lattice tunes over 25 synchrotron periods using a Gaussian bunched beam with $|\Delta\nu_y|=0.05$ injected into the IOTA lattice with no magnet errors. While the vertical rms emittance grows along $Q_x-Q_y=1$, beam loss is only appreciable along $4Q_x=17$. In reality, the field errors in IOTA magnets will excite higher-order betatron resonances, which we will find and compensate for experimentally.\\

The interaction between space-charge fields within bunched beams and the wakefields originating from the vacuum chamber's structure can be understood within a parameter space where the total impedance of the ring and the incoherent tune shift, normalized to the synchrotron tune, serve as independent axes. The structure of this parameter space has been extensively explored in various theoretical models (e.g., \cite{Burov2018, Burov2019, Buffat2021}, all of which predict instability growth or coherent amplification in the strong space-charge regime, in line with experimental findings at the CERN SPS\cite{Bartosik2013}. We propose a comprehensive study to measure the rate of instability growth and head-tail amplification. This study will be conducted as a function of variable wakefields and equilibrium tune shifts using bunched proton beams in IOTA. We will employ a digital wake-building feedback system known as the \textit{waker}\cite{Ainsworth2021} to excite wakefields with arbitrary shapes and magnitudes. Subsequently, we will measure the transverse coherent oscillations as a function of longitudinal position. Concurrently, we will utilize the electron cooler to establish an equilibrium tune shift that can be maintained consistently during each observation. Such an experimental effort will enable us to compare our findings with instability models relevant to high-intensity synchrotrons and storage rings.\\

Mitigation techniques for coherent instabilities \cite{Metral2021} such as TMCI include feedback systems, optics, and RF manipulations, as well as the inclusion of incoherent tune spread using chromaticity and/or dedicated non-linear focusing elements. Non-linear focusing provided by octupole magnets or Gaussian electron lenses leads to chaotic dynamics at large amplitudes, thus increasing beam loss. The demonstration of NIO elements such as the Danilov-Nagaitsev magnets and octupole strings \cite{Danilov2010} to generate variable Landau damping while preserving stable single-particle dynamics is a fundamental goal of the Integrable Optics Test Accelerator. We can verify the preservation of the analytical invariant quantities by measuring turn-by-turn centroid positions of pencil beams kicked at variable amplitudes. Previous experiments on NIO at IOTA using electrons \cite{Valishev2021, Wieland2023} were able to collect data for a limited number of turns, as the chromatic tune spread led to the quick decoherence of the mean position signal of the bunch. Electron cooling can reduce the energy spread of protons, thus reducing the chromatic tune spread, resulting in longer turn-by-turn position datasets. Although the invariance properties of all known NIO systems are violated in the presence of non-linear space-charge forces, we can still measure the effectiveness of these elements in damping instabilities. Using electron cooling and the \textit{waker} system, we can emulate the impedance and the space-charge tune shift of a high-energy ring and then measure the optimum non-linear focusing strength required to damp instabilities with minimum beam loss as a function of ring impedance and space-charge tune shift.\\

\section{Conclusion}
The 2.5~MeV proton program at the Integrable Optics Test Accelerator is designed to explore methods for improving the stability of hadron beams with high phase-space density and total current in synchrotrons and storage rings. The comparatively low-energy regime has the advantage of achieving transverse space-charge tune shifts close to 0.5 with almost zero intrinsic ring impedance, providing a test bench that disentangles space-charge and intensity effects. We will use an electron cooler with cooling times adjustable between 1-10~seconds to compensate for heating mechanisms such as IBS and residual gas scattering. Cooling will enable us to maintain phase-space distributions in equilibrium, and a wake-building feedback system will allow us to generate an artificial ring impedance with arbitrary shape and magnitude. We plan to execute three broad categories of experiments:
(1) Optimizing the IOTA lattice to maximize the space-charge tune shift of the circulating beam while staying within a given emittance growth and beam loss budget. (2) Measuring the dependence of coherent instability growth and head-tail amplification as a function of ring impedance and space-charge tune shift. (3) Demonstrating non-linear integrable optics to minimize instability growth and reduce beam loss in the presence of space-charge. Suitable diagnostics in the proton injector, the IOTA ring, and the electron cooler will equip us to measure beam positions, current, loss, and transverse profiles. Our ongoing work includes finalizing the design of the electron beam transport system, constructing and commissioning the proton injector with the goal of having protons in IOTA in summer 2024, and the first electron cooling in late 2025.\\

\section*{Acknowledgement}
We would like to thank A.~Valishev, A.~Romanov and V.~Lebedev for discussions on proton operations in IOTA. R.~Ainsworth and A.~Burov proposed and developed the \textit{waker} system for instability studies. This manuscript has been authored by Fermi Research Alliance, LLC under Contract No.~DE-AC02-07CH11359 with the U.S.\ Department of Energy, Office of Science, Office of High Energy Physics. This work is also supported by the U.S. Department of Energy, Office of Science, Office of Nuclear Physics under contract DE-AC05-06OR23177. This research is also supported by the University of Chicago.\\

\end{document}